%% file: 0-sample-sigplan.tex
\documentclass[sigplan,10pt]{acmart}
\acmConference[PPoPP'26]{ACM SIGPLAN Annual Symposium on Principles and Practice of Parallel Programming}{January 31--February 04, 2026}{Sydney, Australia}
\acmYear{2026}
\acmISBN{} 
\acmDOI{} 
\startPage{1}
\usepackage{array}
\usepackage{algorithm}
\usepackage{algorithmic}
\usepackage{multirow}
\usepackage{float}
\usepackage{graphicx}
\usepackage{subfig}
\usepackage{ulem}
\usepackage{enumitem}

\AtBeginDocument{%
	}
\renewcommand\footnotetextcopyrightpermission[1]{} 
\begin{document}

\title{\uline{LO}w-c\uline{O}st yet High-\uline{P}erformant \uline{S}parse Matrix-Matrix Multiplication on Arm SME Architectures}

\author{Kelun Lei$^\star$, Hailong Yang$^\star$, Kaige Zhang$^\star$, Kejie Ma$^\star$, Yiqing Wang$^\star$, Xin You$^\star$, Yufan Xu$^\dagger$, Enrique S. Quintana-Orti$^\ddagger$, Zhongzhi Luan$^\star$, Yi Liu$^\star$ and Depei Qian$^\star$}

\affiliation{
  \institution{$^\star$Beihang University, Beijing, China
   $^\dagger$Independent Researcher, Cupertino, USA,
   $^\ddagger$Universitat Politècnica de València
  }
  \country{}
  \city{}
}
\email{{kelunlei, hailong.yang, kaige.zhang, kejiema, YiqingWang, youxin2015, 07680, yi.liu, depeiq}@buaa.edu.cn}
\email{sz.yufanxu@gmail.com, quintana@uji.es}

\fancyhead{}  
\renewcommand\footnotetextcopyrightpermission[1]{} 
\begin{abstract}
Sparse matrix–dense matrix multiplication (SpMM) is a critical kernel in both scientific computing and emerging graph learning workloads. The recent Armv9 architecture introduces Scalable Matrix Extension (SME), enabling tile-based matrix operations with high throughput. However, effectively exploiting both SME and traditional SIMD resources for unstructured sparse workloads remains an open challenge. 
To address this, we propose LOOPS, a hybrid execution framework that combines row-wise CSR-part with vector-wise BCSR-part layout, enabling cooperative utilization of vector instructions (NEON) and Scalable Matrix Extension (SME) resources. LOOPS supports multi-precision SpMM across FP64, FP32, and FP16 via an adaptive two-level parallelization scheme guided by a lightweight performance model. Experimental results on the entire SuiteSparse  on an Apple's M4Pro CPU show that LOOPS achieves average speedups of 9.93$\times$ (FP32)/14.4$\times$ (FP64) against the CPU baseline TACO and 71.3$\times$ (FP32)/54.8$\times$ (FP64) with respect to Armadillo. A comparison of LOOPS running on the same CPU with two GPU methods (cuSPARSE, Magicube) executed on an NVIDIA A100 GPU show average speedups for LOOPS between 19.8$\times$ and 33.5$\times$, depending on the precision. Notably, LOOPS delivers significantly better energy efficiency than the GPU codes on the A100 GPU.
\end{abstract}




\maketitle

\input{1-introduction}
\input{2-background}
\input{3-methodology}

\input{4-evaluation}
\input{5-relatedwork}
\input{6-conclusion}

\bibliographystyle{ACM-Reference-Format}
\makeatletter
\@ACM@balancefalse
\makeatother
\bibliography{sample-base}

\end{document}

%% file: 1-introduction.tex
\section{Introduction}
\label{sec:intro}




As the demand for extreme performance in traditional scientific computing and emerging deep learning applications continues to grow, many hardware vendors have introduced new architectures and instruction sets to accelerate matrix operations. 
Notably, recent high-end CPUs (e.g., Apple M4) incorporate outer-product matrix units, such as Armv9's Scalable Matrix Extension (SME)~\cite{ArmRefManual, ArmSMEGuide}, improving both the efficiency and performance of matrix operations. 
However, while dense matrix operations can efficiently and conveniently leverage such matrix units for acceleration, sparse matrix operations face significant challenges in achieving optimal hardware peak performance. Among sparse matrix operations, Sparse Matrix-Matrix Multiplication (SpMM) has garnered substantial attention for its potential to reduce memory requirements and computation time in various domains, including scientific computing~\cite{anzt2015accelerating,blei2003latent,lan2014sparse}, solvers~\cite{aktulga2014optimizing,knyazev2001toward}, graph neural networks (GNNs)~\cite{kipf2016semi,velivckovic2017graph}, and deep learning~\cite{chien2020adaptive,elsen2020fast,liu2015sparse}. Therefore, investigating how to optimize SpMM to harness matrix units and improve performance on CPUs is of critical importance.

Existing CPU-side SpMM efforts~\cite{rahman2021fusedmm,regnault2023spc5,fu2024jitspmm,zheng2023optimizing,hong2019adaptive} largely focus on vector instruction sets (e.g. NEON and AVX). In addition, code-generation-based sparse compilers ~\cite{kjolstad2017tensor,dias2022sparselnr,chen2018tvm,bansal2023mosaic} and vendor or third-party libraries~\cite{ArmPL2504,sanderson2016armadillo,wang2014intel} also contribute to this space. Therefore, SME-style outer-product matrix units in modern CPUs still remain underutilized in sparse workloads.
In contrast, GPUs with inner-product matrix units (a.k.a, Tensor Core Units, TCUs) have achieved remarkable performance on SpMM tasks. Many works~\cite{okanovic2024high,li2022efficient,zhang2024jigsaw,fan2024dtc,wang2023tc,zhao2025acc,chen2021efficient} have exploited TCUs to accelerate SpMM with specialized block-based sparse formats and computation strategies. Their core optimizations include efficient data movement, overlapping memory loads and computations, pipeline-based decoupling of execution stages, and asynchronous execution. Some studies~\cite{castro2023venom,CuSparse} further explore sparse TCUs for structured sparsity patterns.
However, the above CPU/GPU techniques are not directly applicable to SME-based architectures due to three unique challenges:  
\begin{itemize}[leftmargin=*,align=left]
    \item \textbf{C1: Lack of sparse-specific SME mappings.} SME units are primarily designed for dense GEMM (general matrix-matrix multiplication) kernels and no existing CPU work has established an efficient sparse storage scheme and computation dataflow for it. Inspired by GPU block-based strategies for TCUs, a natural idea is to apply similar blocking for SME. However, the zero-propagation effect in outer-product computation amplifies the cost of padding zeros within sparse blocks, causing severe waste of compute and memory bandwidth.
    
    \item \textbf{C2: Underexplored multi-precision sparse acceleration.} SME natively supports computations in FP64, FP32 and FP16 formats, but existing CPU-side SpMM frameworks seldom optimize for FP16 due to lack of hardware support. As a result, the opportunity to exploit SME for high-performance half-precision sparse computation remains largely unexplored.

    \item \textbf{C3: Balancing vector and matrix unit utilization.} Architectures such as Apple’s M4 series integrate both vector (NEON) and matrix (SME) units, with SME units shared across multiple cores. Our measurements show that assigning all threads exclusively to SME execution often degrades performance for light or irregular workloads due to severe inter-core contention for SME resources, while leaving NEON pipelines underutilized. Since SME instructions execute on a dedicated pipeline distinct from vector instructions, a coordinated strategy that leverages both pipelines can maximize hardware resource utilization.
\end{itemize}

To address these challenges, we present LOOPS, a lightweight and flexible SpMM framework for CPUs with outer-product matrix units. LOOPS adopts a hybrid sparse format combining a CSR-part processed by vector instructions (e.g., NEON), and the vector-wise BCSR part, executed via matrix instructions (e.g., SME). This hybrid design mitigates padding overhead (C1), supports multi-precision SpMM including FP16 (C2), and enables cooperative utilization of vector and matrix units to avoid resource contention (C3). 
At runtime, LOOPS adopts a two-level parallel execution model. At the coarse level, the CSR and BCSR kernels are assigned to disjoint thread groups, enabling inter-kernel parallelism. Within each kernel, fine-grained thread parallelism with a customized implementation is applied to further accelerate computation. A lightweight quadratic performance model, calibrated via representative warm-up runs, predicts the optimal workload partition and thread allocation for a given matrix.

Our evaluation on the entire Suitesparse demonstrates that the LOOPS SpMM implementation delivers significant performance improvements over widely used sparse libraries and compilers. On the Apple M4Pro (12-core) CPU, LOOPS achieves an average speedup of 9.93$\times$ over TACO~\cite{kjolstad2017tensor} and 71.3$\times$ over Armadillo~\cite{sanderson2016armadillo} in FP64 precision. For FP32 precision, LOOPS provides 14.44$\times$ and 54.57$\times$ average speedups over TACO and Armadillo, respectively. For FP16 precision, 
we compare our method against vendor-provided and manual-optimized GPU baselines. On average, LOOPS running on the Apple M4Pro chip outperforms cuSPARSE~\cite{CuSparse} and Magicube~\cite{li2022efficient} executed on an NVIDIA A100 GPU by 33.49$\times$ and 19.79$\times$, respectively. Despite the A100 delivering high throughput in some cases, its power consumption is significantly greater. We observe that LOOPS achieves comparable or superior energy efficiency across various representative matrices. For instance, as detailed in Table~\ref{tab:a100_m4pro}, even when the GPU code executes faster on the A100 GPU, LOOPS' energy efficiency is still up to 20$\times$ higher.

The primary contributions of this paper are as follows:
\begin{itemize}[leftmargin=*,align=left]
    \item We propose LOOPS, a novel hybrid format that partitions the workload between vector and matrix units, mitigating the zero-propagation effect of outer-product computation while preserving data locality.
    
    \item We are the first to exploit SME’s native FP64/FP32/FP16 support for SpMM, enabling high-performance sparse computation among different precisions on CPUs without specialized accelerators.
    
    \item We propose a hardware-aware scheduling strategy with a lightweight performance model that dynamically balances execution between vector and matrix pipelines, alleviating inter-core SME contention and maximizing throughput.

    \item Evaluations on SuiteSparse show LOOPS outperforms all the CPU baselines on M4Pro chip and achieves comparable performance and better energy efficiency against GPU baselines on A100 across FP64, FP32, and FP16 precisions, which demonstrates the effectiveness of the proposed framework and optimizations.
\end{itemize}


%% file: 2-background.tex
\section{Background}
\label{sec:background}

\subsection{SpMM}
\label{sec:background:spmm}
SpMM refers to the multiplication of a sparse matrix $A$ with a dense matrix $B$, producing a dense output matrix $C$. There are two prominent approaches to compute SpMM: inner-product based and outer-product based methods.

\textbf{Inner-Product based SpMM:} This approach computes each element $c_{ij}$ in the output matrix $C$ as the dot (inner) product of the 
$i$-th row of $A$ and the $j$-th column of $B$ 
as $c_{ij} = \sum_k{a_{ik} \cdot b_{kj}}$%
, where $a_{ik}$ and $b_{kj}$ are non-zero elements from $A$ and $B$, respectively. This method iterates over rows of $A$ and columns of $B$, requiring efficient sparse row traversal and indirect memory accesses to $B$. While this approach aligns well with the row-major storage of sparse matrices (e.g., CSR format), its efficiency is often constrained by irregular memory access patterns and limited opportunities to leverage data reuse in $B$.

\textbf{Outer-Product based SpMM:} This alternative computes $C$ by iterating over the non-zero elements of A and accumulating rank-1 updates as $C = \sum_k{a_{:,k} \cdot b_{k,:}}$%
, where $a_{:,k}$ is the sparse $k$-th column of $A$, and $b_{k,:}$ is the dense $k$-th row of $B$. This method emphasizes column-wise sparsity and allows for better data reuse of $b_{k,:}$. Moreover, it is naturally suited to hardware accelerators equipped with outer-product matrix units, as each rank-1 update can be mapped to efficient outer-product computations. However, the outer-product method requires carefully orchestrated memory and compute strategies to avoid excessive intermediate data accumulation and maintain computational efficiency. 

\subsection{Modern multi-Core CPUs and matrix extension}
\label{sec:background:sme}

\begin{table}[htbp]
  \centering
  \small
  \caption{Peak performance comparison (TFLOPS for floating point, TOPS for INT8) of Apple M4Pro (12-core CPU) and NVIDIA A100 (with Tensor Cores).}
  \label{tab:peak_performance}
  \begin{tabular}{lcccc}
    \toprule
    \multirow{2}{*}{\textbf{Device}} & \multicolumn{4}{c}{\textbf{Precision (TFLOPS/TOPS)}} \\
    \cmidrule(lr){2-5}
     & \textbf{FP64} & \textbf{FP32 / TF32} & \textbf{FP16} & \textbf{INT8} \\
    \midrule
    Apple M4Pro  & 1.0  & 4.1 (FP32)   & 8.3   & 33.3 \\
    NVIDIA A100  & 19.5 & 156 (TF32)   & 312   & 624 \\
    \bottomrule
  \end{tabular}
\end{table}

The advanced Armv9 architecture introduces Scalable Matrix Extension (SME) which enables innovation and acceleration across GEMM-like workloads, including those arising in AI (artificial intelligence) and machine learning applications. Although the theoretical peak performance of GPU typically surpasses that of SME-equipped CPUs as shown in Table~\ref{tab:peak_performance}, the gap can narrow significantly under real-world constraints such as memory bandwidth, workload sparsity, and power consumption. 
Consider the Apple M4Pro CPU, for example. Compared to high-end professional GPUs, the M4Pro chip offers several compelling advantages. First, the M4Pro demonstrates substantially lower power consumption, which not only reduces operational costs but also alleviates thermal design constraints. Specifically, the Thermal Design Power (TDP) of M4Pro is 40W which is significantly lower than 250W of A100. Despite its lower power envelope, the M4Pro achieves attainable computational performance, particularly in low-precision arithmetic operations (e.g., INT8 or FP16), where it exhibits a high throughput. This combination of low cost, energy efficiency, and strong low-precision performance results in a markedly higher performance-per-watt ratio compared to traditional GPUs, making the M4Pro an efficient and scalable solution for edge computing and AI inference workloads.

%% file: 3-methodology.tex
\section{Methodology}
\label{sec:method}




\subsection{Overview}
\label{sec:method:overview}
\begin{figure*}[t]
	\centering
	\scriptsize
	\includegraphics[width=\linewidth]{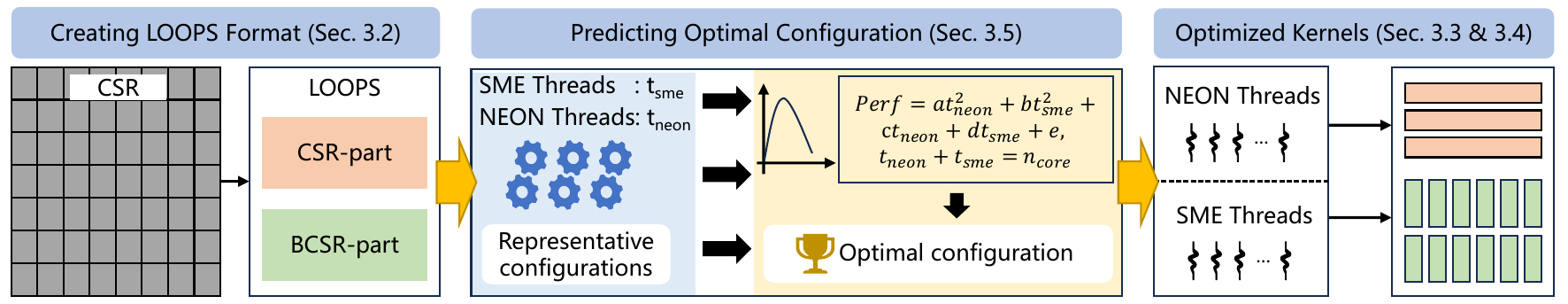}
	\caption{The overview of our SpMM pipeline.}
	\label{fig:overview}
\end{figure*}
Figure~\ref{fig:overview} presents an overview of our method. First, to address \textbf{C1}, LOOPS employs a hybrid format that partitions the sparse matrix into a CSR-part and a BCSR-part in a row splitting way. The BCSR-part is carefully structured to align with SME’s outer-product model without excessive padding. To determine three key parameters (\textbf{C3}), namely 1) the row boundary $r_{\text{boundary}}$ that separates the CSR and BCSR regions, 2) the number of threads $t_{\text{neon}}$ assigned to execute the SpMM of the CSR part, and 3) the number of threads $t_{\text{sme}}$ for the BCSR part, we first fit a performance model using a representative set of parameter configurations to predict the optimal thread setup. Then, the row boundary $r_{\text{boundary}}$ serves as a balancing factor that equalizes the workload and computational capability between the NEON and SME units. This parameter is set by solving the following equation
\begin{equation}
\label{eq:rb}
r_{\text{boundary}} \cdot \text{TP}_{\text{neon}} \cdot t_{\text{neon}} = (r_{\text{total}} - r_{\text{boundary}}) \cdot \text{TP}_{\text{sme}} \cdot t_{\text{sme}},
\end{equation}
where $r_{\text{total}}$ is the total number of rows, while $\text{TP}_{\text{neon}}$ and $\text{TP}_{\text{sme}}$ specify the throughput of two processing units. 
Finally, the matrix is passed to the execution stage, where our custom high-performance kernel performs the matrix multiplication. To resolve \textbf{C2}, LOOPS provides dedicated support for FP64, FP32 and FP16. In particular, our FP16 BCSR-part kernel exploits SME’s 2-way \texttt{fmopa} instructions, enabling high-performance mixed-precision execution. 

\subsection{The LOOPS format}
\label{sec:method:loops}



\subsubsection{Data layout representation}
\label{sec:method:loops:format}
On architectures whe\-re multiple cores share SME units, resource contention among the threads can degrade performance. To address this, we propose a hybrid sparse format, LOOPS, which combines a row-wise \textit{CSR-part} layout with a vector-wise \textit{BCSR-part} layout to leverage the benefits of both NEON and SME hardware resources. This hybrid strategy allows one portion of the matrix to be handled efficiently with NEON vector instructions due to the row-aligned memory layout of CSR part. Meanwhile, the other portion benefits from SME's matrix-level parallelism through block outer-products on structured BCSR blocks. The block mapping also enables spatial reuse, making it highly cache- and compute-efficient.


While block formats such as BCSR offer high throughput under regular sparsity patterns, they introduce significant overhead for highly sparse matrices due to the increased number of padding zeros within blocks as well as severe useless computations. 
Instead of using 2D tiles, LOOPS employs an asymmetric tile shape of $\texttt{vector\_size} \times 1$. This design is compatible with the outer-product computation model of SME, in which each compute unit operates on a column tile from the sparse matrix $A$ and the corresponding row tile from the dense matrix $B$. The use of this narrow tile shape reduces padding overhead and allows the tile to be used directly as input to the SME outer-product unit, leading to improved performance and resource efficiency.




\subsubsection{Conversion from CSR to LOOPS}
\label{sec:method:loops:conversion}

The transformation from a conventional CSR format to our LOOPS format involves two distinct stages as shown in Algorithm~\ref{algo:conversion}: the construction of a CSR-part segment for the top $r_{boundary}$ rows of the sparse matrix and the construction of a BCSR-part segment for the remaining rows.

\textbf{Step 1: Constructing CSR-part.}
Given a standard CSR matrix $(\texttt{row\_ptr}, \texttt{col\_idx}, \texttt{vals})$, we extract the top $r_{boundary}$ rows and form a CSR-part layout. This step involves copying relevant data from the original arrays, preserving the row structure. The new arrays $\texttt{row\_ptr'}$, $\texttt{col\_idx'}$, and $\texttt{vals'}$ are updated accordingly to reflect the reduced row span.

\textbf{Step 2: Constructing BCSR-part.}
For the remaining rows, we employ a tile-based layout. We define fixed tile dimensions $(B_r, B_c)$ and iterate over each nonzero element to compute its corresponding tile location $(\texttt{tile\_r}, \texttt{tile\_c})$ and intra-tile offset. A hash map is maintained to record each unique tile’s address and to group the values accordingly.


\begin{algorithm}[htbp]
\caption{Conversion from CSR to LOOPS Format}
\label{algo:conversion}
\begin{algorithmic}[1]
\STATE \textbf{Input:} CSR matrix \texttt{row\_ptr}, \texttt{col\_idx}, \texttt{vals}; row threshold $r_{boundary}$; row count $row_{end}$; tile size $(B_r, B_c)$
\STATE \textbf{Output:} LOOPS format: CSR-part and BCSR-part

\STATE Initialize \texttt{row\_ptr'}, \texttt{col\_idx'}, \texttt{vals'} $\leftarrow$ empty arrays
\STATE Initialize \texttt{tile\_map} $\leftarrow$ empty map

\STATE // Step 1: Build CSR-part
\FOR{$i \gets 0$ to $r_{boundary}-1$}
    \FOR{$k \gets \texttt{row\_ptr}[i]$ to $\texttt{row\_ptr}[i+1] - 1$}
        \STATE Append \texttt{col\_idx[$k$]} to \texttt{col\_idx'}
        \STATE Append \texttt{vals[$k$]} to \texttt{vals'}
    \ENDFOR
    \STATE Append current length of \texttt{vals'} to \texttt{row\_ptr'}
\ENDFOR

\STATE // Step 2: Build BCSR-part
\FOR{$i \gets r_{boundary}$ to $r_{end}-1$}
    \FOR{$k \gets \texttt{row\_ptr}[i]$ to $\texttt{row\_ptr}[i+1] - 1$}
        \STATE $j \gets \texttt{col\_idx}[k]$
        \STATE $tile_r \gets i \div B_r$,\quad $tile_c \gets j \div B_c$
        \STATE $offset \gets (i \bmod B_r) \cdot B_c + (j \bmod B_c)$
        \IF{$(tile_r, tile_c) \notin \texttt{tile\_map}$}
            \STATE $\texttt{tile\_map}[(tile_r, tile_c)] \gets$ new tile
        \ENDIF
        \STATE $\texttt{tile\_map}[(tile_r, tile_c)][offset] \gets \texttt{vals}[k]$
    \ENDFOR
\ENDFOR
\end{algorithmic}
\end{algorithm}


\subsection{Kernel design}
\label{sec:method:kernel}
\begin{figure}[htbp]
	\centering
	\scriptsize
	\includegraphics[width=\linewidth]{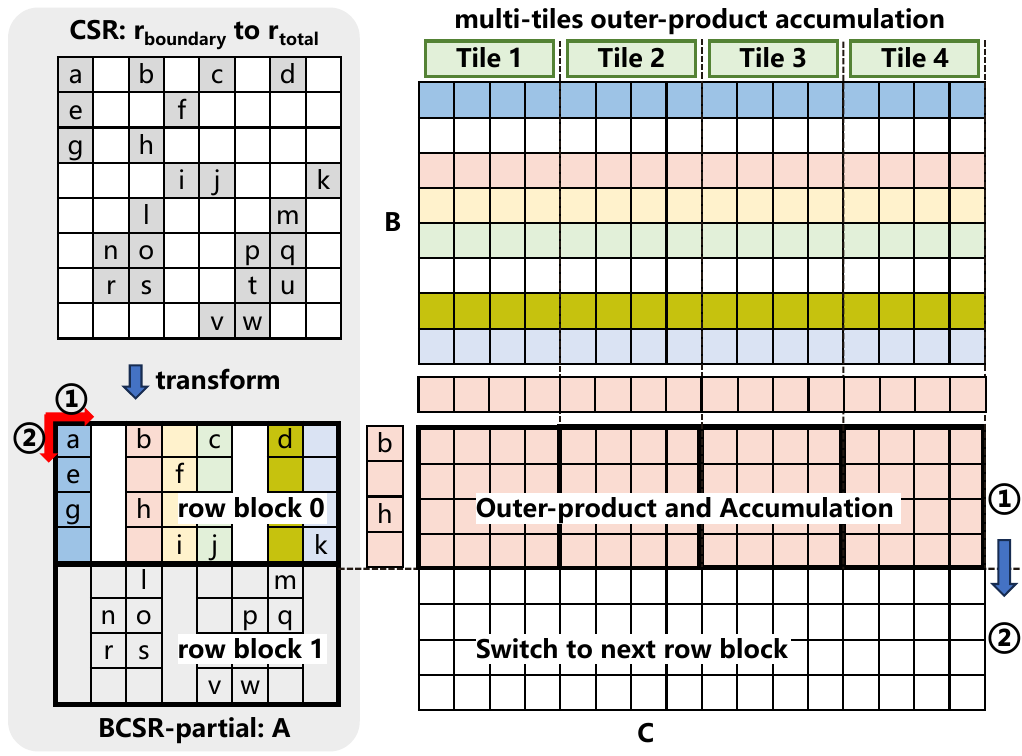}
	\caption{The BCSR-part SpMM workflow.}
	\label{fig:multi-tile-workflow}
\end{figure}
\begin{algorithm}[htbp]
\caption{BCSR-part SpMM with SME (FP64)}
\label{algo:kernel-fp64}
\begin{algorithmic}[1]
\STATE \textbf{Input:} BCSR matrix $row\_ptr$, $col\_idx$, $vals$; row count $n_r$; row block count $row_b$; tile count $n_b$; tile row size $B_r$; tile column size $B_c$; N dim $N$; dense matrix $B$
\STATE \textbf{Output:} dense matrix $C$

\STATE $\textit{cntd} \gets vector\_length \div sizeof(double)$

\FOR{$i \gets 0$ to $row_b$}
    \FOR{$j \gets 0$ to $N$ by $\textit{cntd}$}
        \STATE $r_{begin} \gets i \times B_r$
        \FOR{$cnt \gets 0$ to $\textit{cntd}$}
            \IF{$cnt + r_{begin} \geq n_r$}
                \STATE \textbf{break}
            \ENDIF
            \STATE load \&$C$[$(r_{begin} + cnt) * N + j$] to ZA tile
        \ENDFOR

        \FOR{$k \gets row\_ptr[i]$ to $row\_ptr[i+1]$}
            \STATE $col\_i \gets col\_idx[k] \times B_c$
            \STATE $val\_offset \gets k \times (B_r \times B_c)$
            \STATE $r_A \gets vload(\&vals[val\_offset])$
            \STATE $r_B \gets vload(\&B[col\_i \times N + j])$

            \STATE // Floating-point outer product and accumulate
            \STATE $fmopa(r_A, r_B)$
        \ENDFOR

        \FOR{$cnt \gets 0$ to $\textit{cntd}$}
            \IF{$cnt + r_{begin} \geq n_r$}
                \STATE \textbf{break}
            \ENDIF
            \STATE store ZA tile to \&$C$[$(r_{begin} + cnt) * N + j$]
        \ENDFOR
    \ENDFOR
\ENDFOR
\end{algorithmic}
\end{algorithm}

\textbf{AXPY based NEON kernel for CSR part.}
For the CSR-part matrix $A_{csr}$, we design a NEON-optimized kernel that leverages row-wise sparsity and SIMD vectorization. The input sparse matrix is stored in a standard CSR layout, and the kernel performs row-wise accumulation by iterating over each nonzero element in the row of $A$ and applying an AXPY operation with the corresponding row of matrix $B$. Specifically, for each nonzero element with column index $j$, we load the contiguous segments of $B$ in row $j$ into NEON registers and use fused multiply-add instructions to scale and accumulate these segments into the output row. 

\textbf{Outer-product based SME kernel for BCSR part.}
To preserve output locality and enhance cache efficiency, we do not compute the full outer product of an entire column of BCSR-part matrix $A_{bcsr}$ and a full row of matrix $B$ to form the partial result of the entire matrix $C$. Instead, the matrix $A_{bcsr}$ is divided into a series of row blocks, each containing multiple rows of $A_{bcsr}$ and further organized into a set of $B_r \times B_c$-shaped tiles. Consider the FP64 case first. In our setting, we fix $B_r=\textit{cntd}$ and $B_c=1$, where $\textit{cntd}$ equals the maximum number of FP64 elements that a vector register can accommodate.
The naive implementation of the FP64 SME-based BCSR-part SpMM is illustrated in Algorithm~\ref{algo:kernel-fp64}. The outermost loop (Line~4) iterates over all row blocks of matrix $A_{bcsr}$; the second loop (Line~5) sweeps across the columns of matrix $B$ in tiles of size $\texttt{cntd}$. These two loops together compute a $\texttt{cntd} \times \texttt{cntd}$ tile of the output matrix $C$. Lines~7--12 handle the loading of this output tile into a \texttt{ZA} tile register for accumulation. The core computation takes place in Lines~13--20, which iterate over all $\textit{cntd} \times 1$ tiles within the current row block of $A_{bcsr}$, and the corresponding $1 \times \textit{cntd}$ tiles in $B$. Each valid pair contributes to the outer-product update via SME instructions \texttt{fmopa}. Finally, Lines~21--26 store the accumulated result from the \texttt{ZA} tile back to memory.
To further saturate the available \texttt{ZA} tile bandwidth and exploit the parallelism offered by the SME architecture, we implement a \textit{multi-tile outer-product accumulation} strategy. As shown in Figure~\ref{fig:multi-tile-workflow}, instead of loading a single $1 \times \textit{cntd}$ tile of $B$ at a time, we load multiple tiles of $B$ simultaneously. This enables issuing multiple \texttt{fmopa} instructions in parallel, each utilizing a different \texttt{ZA} tile. This optimization effectively increases the throughput of the innermost computation loop (Lines~13--20 in Algorithm~\ref{algo:kernel-fp64}).

\begin{figure}[htbp]
	\centering
	\scriptsize
	\includegraphics[width=\linewidth]{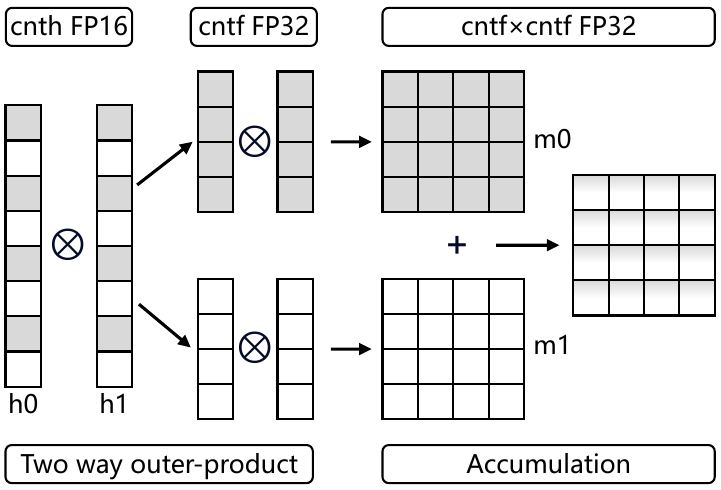}
	\caption{Illustration of FP16 two way \texttt{fmopa} instruction.}
	\label{fig:two-way-fmopa}
\end{figure}

\begin{algorithm}[htbp]
\caption{Core computation loop of BCSR-part SpMM with SME (FP16)}
\label{algo:kernel-fp16}
\begin{algorithmic}[1]
\STATE \textbf{Input:} BCSR matrix $row\_ptr$, $col\_idx$, $vals$; row block index $i$; block row size $B_r$; block column size $B_c$; N dim $N$; dense matrix $B$; column index of B $j$

\STATE // Unroll the loop by 2 for two way outer-product.
\FOR{$k \gets row\_ptr[i]$ to $row\_ptr[i+1]-1$ by $2$}
    \STATE $col\_i0 \gets col\_idx[k] \times B_c$
    \STATE $col\_i1 \gets col\_idx[k+1] \times B_c$
    \STATE $val\_offset0 \gets k \times (B_r \times B_c)$
    \STATE $val\_offset1 \gets (k+1) \times (B_r \times B_c)$
    
    \STATE $tmp0 \gets vload(\&vals[val\_offset0])$
    \STATE $tmp1 \gets vload(\&vals[val\_offset1])$
    \STATE $r0_{A} \gets vzip1(tmp0, tmp1)$
    \STATE $r1_{A} \gets vzip2(tmp0, tmp1)$

    \STATE $tmp0 \gets vload(\&B[col\_i0 \times N + j])$
    \STATE $tmp1 \gets vload(\&B[col\_i1 \times N + j])$
    \STATE $r0_B \gets vzip1(tmp0, tmp1)$
    \STATE $r1_B \gets vzip2(tmp0, tmp1)$

    \STATE // Four ZA tiles Floating-point outer product and accumulate (widening, 2-way, FP16 to FP32)
    \STATE $fmopa(r0_A, r0_B)$
    \STATE $fmopa(r0_A, r1_B)$
    \STATE $fmopa(r1_A, r0_B)$
    \STATE $fmopa(r1_A, r1_B)$
\ENDFOR

\STATE // A possible remaining tile is paired with the all-zero register and performs the same computation.
\end{algorithmic}
\end{algorithm}
For FP32, the only difference is that we fix $B_r=\textit{cntf}$, where $\textit{cntf}$ denotes the maximum number of FP32 elements that can be held in a single vector register. However, for FP16, since the f16f16f32 computation mode is the common precision setting in real-world applications and is supported by many vendor-provided libraries~\cite{ArmPL2504,wang2014intel}, we reorganize the computation loops to align with the vectorized input-output format and accumulation constraints. 
Specifically, we utilize the FP16 sum of outer products and accumulate instruction which performs 2-way outer products accumulating the result in FP32 ZA tiles. As shown in Figure~\ref{fig:two-way-fmopa}, this instruction takes two registers h0 and h1 of type FP16, each register holds $cnth$ FP16 elements. Then, the $\textit{cntf}=cnth \div 2$ elements with indices $2i$ of h0 and h1 are widened to FP32 and an f32f32f32 outer-product is performed to obtain a $\textit{cntf} \times \textit{cntf}$ matrix $m0$. Similarly, the $\textit{cntf}$ elements with indices $2i+1$ participate in an analogous operation to obtain a matrix $m1$ of the same size. Finally, $m0$ and $m1$ are element-wise accumulated to obtain the final FP32 result matrix.
Leveraging this instruction, we design and implement the FP16 BCSR-part kernel in Algorithm~\ref{algo:kernel-fp16}. To accumulate the result to the appropriate place of result matrix $C$, we need to reorganize the elements within the vector register before computation. Specifically, as shown in Algorithm~\ref{algo:kernel-fp16}, the loop in Line 3 processes two $cnth \times 1$ tiles (Lines 8--9) within the same row block. Then, Lines 10--11 employ the \texttt{vzip1} and \texttt{vzip2} instructions to perform register shuffling. Specifically, these instructions interleave the lower and higher halves of the loaded data, respectively, to form the correct order required for subsequent computation. For the possible remaining tile, we pair it with zero-filled vector register to ensure alignment and correctness of the computation pipeline.



\subsection{Parallelization}
\label{sec:method:parallel}
To concurrently process the CSR-part and BCSR-part regions of the sparse matrix, we create two thread groups via OMP parallel sections to assign independent computation paths to different parts of the matrix. The first section handles the CSR part with a NEON kernel. Simultaneously, the second section processes the BCSR part using an SME kernel.
This OpenMP-based division enables concurrent utilization of both NEON and SME execution pipelines, thereby maximizing throughput. 
Moreover, LOOPS is designed to facilitate thread-level parallelism without incurring synchronization overhead. Specifically, the CSR-part region is partitioned row-wise, and the BCSR-part region is partitioned row-block-wise, allowing independent threads to operate on disjoint subsets of rows or row blocks. Since each output row in the result matrix $C$ is exclusively assigned to a single thread, inter-thread write conflicts are inherently avoided, eliminating the need for explicit atomics or locks.
Furthermore, the number of threads for each parallel section can be tuned independently, allowing further flexibility in hardware resource allocation.
This hybrid and hierarchical approach to parallelism, enabled by the LOOPS format, delivers both compute efficiency and scalability across NEON and SME units.


\subsection{Adaptive scheduling for heterogeneous execution}
\label{sec:method:schedule}

\subsubsection{Problem formulation}
\label{sec:method:schedule:problem}

Modern CPUs often contain diverse execution units such as different SIMD extensions and dedicated co-processors, which use distinct pipelines. Therefore, to efficiently utilize these heterogeneous units, distributing workloads across them is critical to achieving high performance. Take M4Pro chip as an example. The scheduling problem is to determine the optimal thread partitioning between SME and NEON units for a given workload such that the predicted execution performance is maximized. Given a fixed total number of threads $T$, the search space consists of all $(t_{\text{neon}}, t_{\text{sme}})$ combinations satisfying $t_{\text{neon}} + t_{\text{sme}} \le T$.

\subsubsection{Performance prediction model}
\label{sec:method:schedule:model}

To model the performance impact of different thread partitions, we fit a polynomial regression model based on empirical measurements from a set of candidate configurations. The performance model is a quadratic function without cross terms since NEON and SME execute in their dedicated pipelines:
\begin{equation}
\label{eq:perf-model}
\text{perf}(x, y) = a_0 + a_1 x + a_2 y + a_3 x^2 + a_4 y^2,
\end{equation}
where $x$ and $y$ represent the number of threads assigned to NEON and SME units respectively, and $\text{perf}(x, y)$ is the predicted performance score. Coefficients $(a_0, \ldots, a_4)$ are computed using least squares fitting over the candidate set. The fitted model provides a lightweight analytical cost model to guide scheduling decisions at runtime. 

\subsubsection{Runtime scheduling strategy}
\label{sec:method:schedule:strategy}

At runtime, we enumerate all valid combinations of thread assignments that satisfy $t_{\text{neon}} + t_{\text{sme}} \le T$, where $T$ is the total number of cores. This enumeration is fast due to the limited core count. For each configuration, we compute the predicted performance using Equation~(\ref{eq:perf-model}), selecting the configuration with the highest predicted performance:
\begin{equation}
(t_{\text{neon}}^*, t_{\text{sme}}^*) = \arg\max_{x + y \le T} \text{perf}(x, y).
\end{equation}

This strategy enables the system to dynamically balance the compute workload across NEON and SME units, leveraging the strengths of both execution pipelines. Moreover, this scheduling mechanism is compatible with other adaptive techniques such as tiling and micro-kernel selection, and can be easily integrated into a larger auto-tuning framework.

%% file: 4-evaluation.tex
\section{Evaluation}
\label{sec:evaluation}

\subsection{Evaluation setup}
\label{sec:evaluation:setUp}
\textbf{Hardware platform.} Experiments are conducted on an Apple Mac Mini equipped with the M4Pro chip. 
To the best of our knowledge, Apple M4Pro is the first publicly released commodity processor which adopts the Armv9 architecture. The chip features 12 CPU cores sharing two SME units. The system runs macOS 15.3.1 and all CPU codes are compiled using Clang 20.1.2 with \texttt{-O3}, \texttt{-fopenmp}, and \texttt{-march=armv9-a+ sme+sve} options enabled. GPU experiments are conducted on a platform equipped with Intel Xeon Gold 6336Y (ICELAKE) CPU and NVIDIA A100-PCIE-40GB GPU. The system of the GPU platform runs Ubuntu 22.04.3 LTS and all the GPU methods compiled using cuda-12.3 nvcc (driver version 555.42.02) with the \texttt{-O3} and \texttt{-gencode arch=compute\_80, code=sm\_80} options on.

\textbf{Baselines.} We compare our LOOPS SpMM kernel with the following baselines:

\begin{itemize}[leftmargin=*,align=left]
  \item \textbf{TACO}~\cite{kjolstad2017tensor}: A tensor algebra compiler that generates sparse code based on user-defined format and loop scheduling.
  \item \textbf{Armadillo}~\cite{sanderson2016armadillo}: A high-level linear algebra library for C++ that provides sparse matrix support.
  \item \textbf{cuSparse}~\cite{CuSparse}: NVIDIA’s highly-optimized sparse linear algebra library. 
  \item \textbf{Magicube}~\cite{li2022efficient}: A state-of-the-art GPU SpMM approach that uses tiling and memory coalescing for high-throughput tensor core execution. We use their open-source implementation.
\end{itemize}

TACO and Armadillo are CPU codes evaluated under FP64 and FP32 precision settings; they do not support FP16. For the GPU codes, cuSparse is evaluated using all three precisions while Magicube only supports FP16. We fix \( N = 32 \) (the number of columns in dense matrix $B$) throughout all experiments, which reflects the common use case in deep learning and scientific kernels~\cite{kipf2016semi,gerogiannis2023spade,fu2024jitspmm}.

\textbf{Datasets.} We evaluate our method on the entire SuiteSparse Matrix Collection~\cite{davis2011university}, comprising 4822 sparse matrices that span diverse domains and vary significantly in terms of size, sparsity pattern, and block density. This diversity ensures a comprehensive assessment of our method across representative sparse workloads. Meanwhile, we further select 20 matrices widely tested in previous work~\cite{lu2023dasp} in Table~\ref{tab:detailed} to analyze the advantages of LOOPS in more detail. These group of sparse matrices come from multiple application domains and their scale ranges 0.8M to 59.5M non-zero values, and 36K to 5.6M rows.

\begin{table}
\footnotesize
\caption{Representative matrices from the SuiteSparse Matrix Collection with feature values. $nrow$ denotes the matrix row number. $NNZ$ denotes the nonzero element number. $NNZ_{x}$, $NNZ_{mean}$, and $NNZ_{std}$ denote the (maximum/minimum), mean, and standard deviation of the non-zero element number per row.}
\label{tab:detailed}
\begin{tabular}{llrrlrr}
    \toprule
    Id & Matrix & nrow & NNZ & NNZ$_{x}$ & NNZ$_{mean}$ & NNZ$_{std}$ \\
    \midrule
    m1 & circuit5M & 5.6M & 59.5M & 1.3M/1 & 10.71 & 1356.62 \\
    m2 & Si41Ge41H72 & 0.2M & 15.0M & 662/13 & 80.86 & 126.97 \\
    m3 & Ga41As41H72 & 0.3M & 18.5M & 702/18 & 68.96 & 105.39 \\
    m4 & in-2004 & 1.4M & 16.9M & 7753/0 & 12.23 & 37.23 \\
    m5 & eu-2005 & 0.9M & 19.2M & 6985/0 & 22.30 & 29.33 \\
    m6 & pwtk & 0.2M & 11.6M & 180/2 & 53.39 & 4.74 \\
    m7 & FullChip & 3.0M & 26.6M & 2.3M/1 & 8.91 & 1806.80 \\
    m8 & mip1 & 0.1M & 10.4M & 66K/4 & 155.77 & 350.74 \\
    m9 & mc2depi & 0.5M & 2.1M & 4/2 & 3.99 & 0.08 \\
    m10 & webbase-1M & 1.0M & 3.1M & 4700/1 & 3.11 & 25.35 \\
    m11 & shipsec1 & 0.1M & 7.8M & 102/24 & 55.46 & 11.07 \\
    m12 & econ\_fwd500 & 0.2M & 1.3M & 44/1 & 6.17 & 4.44 \\
    m13 & scircuit & 0.2M & 1.0M & 353/1 & 5.61 & 4.39 \\
    m14 & pdb1HYS & 0.0M & 4.3M & 204/18 & 119.31 & 31.86 \\
    m15 & consph & 0.1M & 6.0M & 81/1 & 72.13 & 19.08 \\
    m16 & cant & 0.1M & 4.0M & 78/1 & 64.17 & 14.06 \\
    m17 & cop20k\_A & 0.1M & 2.6M & 81/0 & 21.65 & 13.79 \\
    m18 & dc2 & 0.1M & 0.8M & 114K/1 & 6.56 & 361.50 \\
    m19 & rma10 & 0.0M & 2.4M & 145/4 & 50.69 & 27.78 \\
    m20 & ASIC\_680k & 0.7M & 3.9M & 395K/1 & 5.67 & 659.81 \\
    \bottomrule
\end{tabular}
\end{table}

\textbf{Metrics.} We report end-to-end SpMM compute throughput (in billions of floating-point operations, or GFLOPS), averaged over 1000 runs for each matrix. For the GPU codes, the host-device transfer time is excluded. For the CPU codes, only execution time is measured, excluding format transformation overhead.

\subsection{Overall evaluation}
\label{sec:evaluation:overall}
\begin{figure}[htbp]
	\centering
    \subfloat[FP64]
    {
	\includegraphics[width=\linewidth]{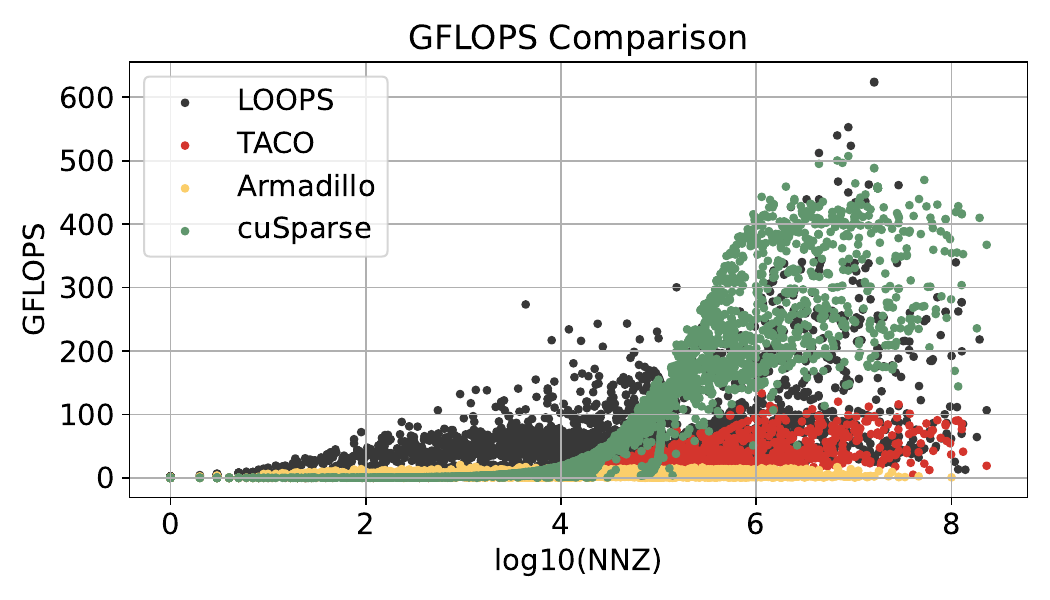}
    }
    
    \subfloat[FP32]
    {
	\includegraphics[width=\linewidth]{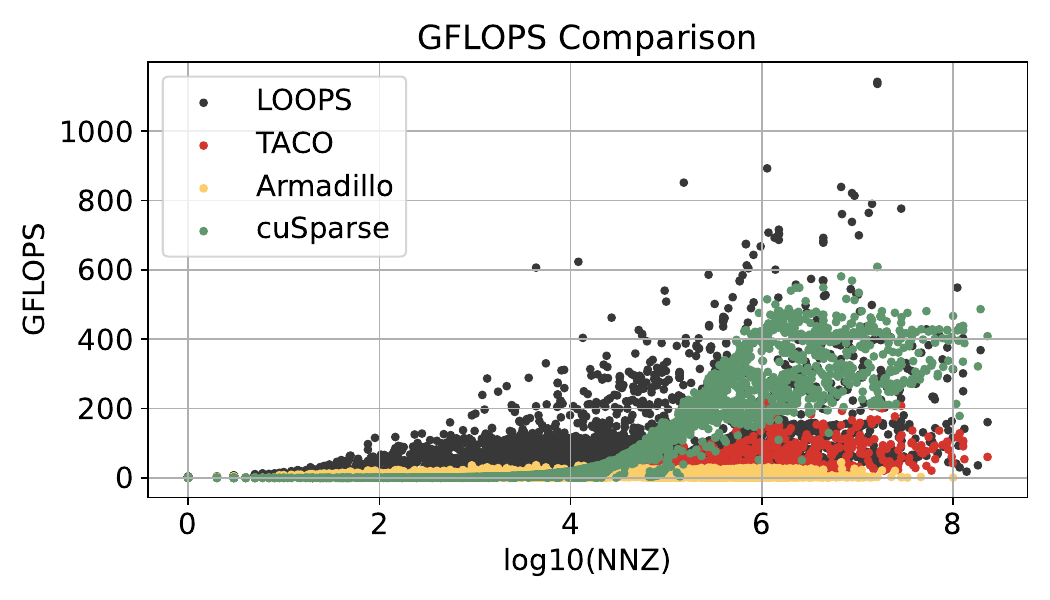}
    }
    
    \caption{Overall SpMM performance (in GFLOPS) of CPU- (on Apple M4Pro) and GPU SpMM methods (on NVIDIA A100) in (a) FP64 and (b) FP32 precisions.}
	\label{fig:gflops-fp64-fp32}
\end{figure}

\begin{figure}[htbp]
    \centering

    \includegraphics[width=\linewidth]{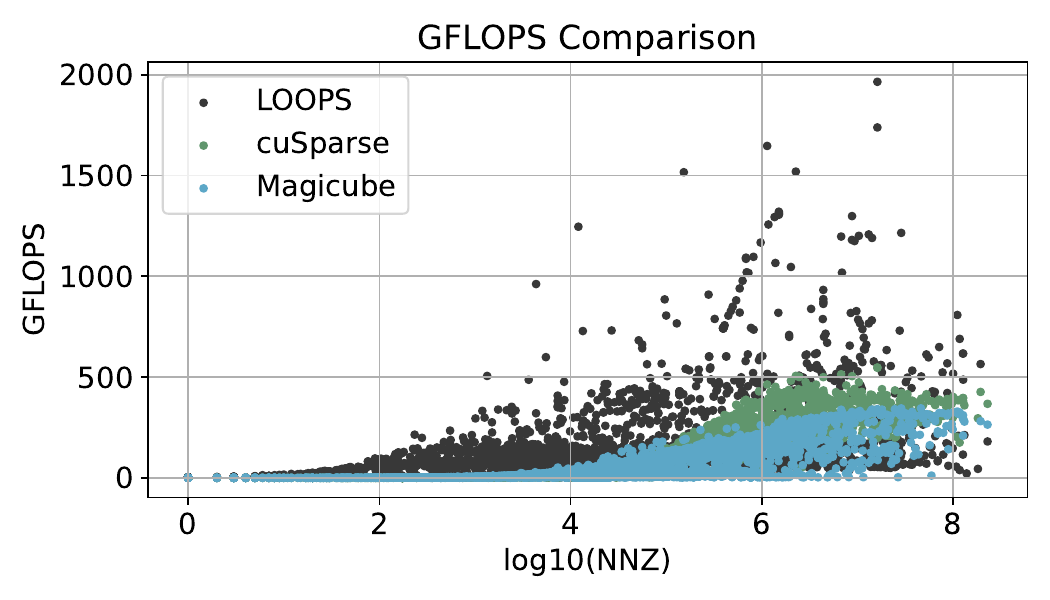}
    
    \caption{Overall SpMM performance (in GFLOPS) comparison of our method (on Apple M4Pro) with GPU SpMM methods (on NVIDIA A100) in FP16 precision.}
    \label{fig:gflops-fp16}
\end{figure}

\begin{figure*}[htbp]
	\centering
    \subfloat[FP64]
    {
	\includegraphics[width=\linewidth]{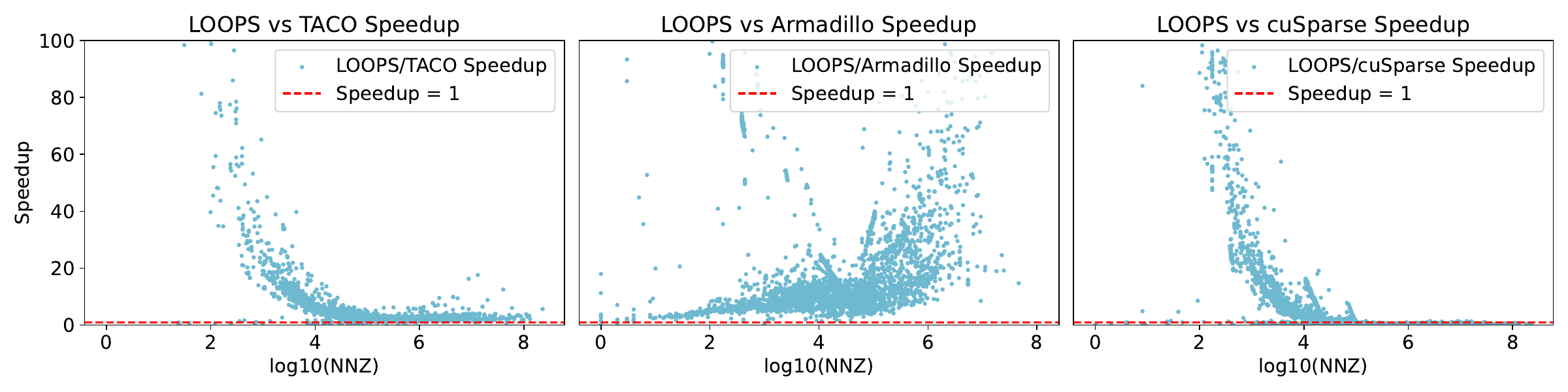}
    }
    
    \subfloat[FP32]
    {
	\includegraphics[width=\linewidth]{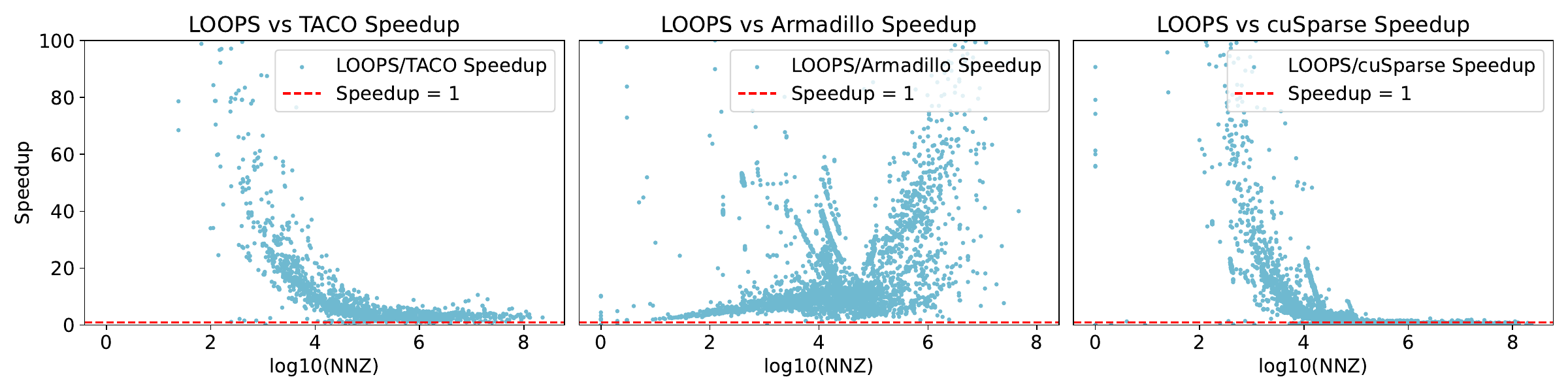}
    }

    \subfloat[FP16]
    {
        \includegraphics[width=\linewidth]{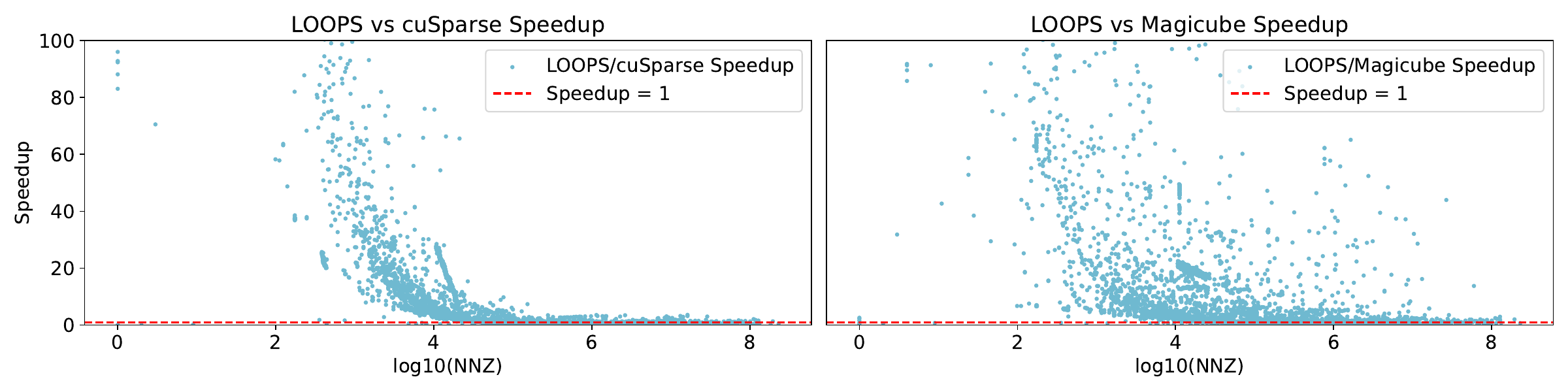}
    }
    
    \caption{Performance speedup of our work, LOOPS, relative to the CPU- (TACO, Armadillo on Apple M4Pro) and GPU SpMM works (cuSparse, Magicube on NVIDIA A100) in (a) FP64, (b) FP32, and (c) FP16 precisions.}
    \label{fig:speedups-fp64-fp32}
\end{figure*}

Figure~\ref{fig:gflops-fp64-fp32} depicts the GFLOPS performance of the CPU methods and cuSparse, ordered by increasing NNZ count (in log scale), in FP64 and FP32 precisions. Figure~\ref{fig:gflops-fp16} illustrates the GFLOPS performance of LOOPS, cuSparse, and Magicube in FP16 precision (TACO and Armadillo are not included as they do not support FP16 precision). 
In addition, we present the speedup ratios of LOOPS relative to each baseline in Figure~\ref{fig:speedups-fp64-fp32}. 

On the Apple M4Pro processor, LOOPS outperforms all CPU baselines across most matrices in all tested precisions. Specifically, compared to TACO, LOOPS delivers higher performance on 96.8\% and 99.0\% of the datasets, with an average speedup of 9.93$\times$ and 14.4$\times$ in FP64 and FP32 precisions, respectively. Furthermore, LOOPS outperforms Armadillo on 99.6\% and 99.7\% of the datasets, with an average speedup of 71.3$\times$ and 54.8$\times$ in FP64 and FP32 precisions, respectively. 
LOOPS also shows competitive performance with GPU baselines. Specifically, compared to cuSparse, LOOPS attains higher performance on 59.5\%, 69.3\%, and 75.8\% of the datasets, with an average speedup of 21.8$\times$, 28.3$\times$, and 33.5$\times$ in FP64, FP32, and FP16 precisions, respectively. Finally LOOPS improves Magicube on 84.1\% of the datasets, with an average speedup of 19.8$\times$ in FP16 precision. 

Experimental results demonstrate that LOOPS achieves significantly higher throughput by effectively leveraging the matrix-level parallelism offered by SME and the vector-level capabilities of NEON. Compared to GPU approaches, LOOPS delivers superior performance when the NNZ is smaller than $10^6$, as GPU methods often suffer from under-utilization due to insufficient workload to saturate their compute and memory resources in those cases. 
For instance, LOOPS attains more than $30\times$ speedup on matrices such as \texttt{wheel\_4\_1}, \texttt{lp\_afiro}, and the \texttt{bibd\_9\_5} series, all of which contain fewer than 2000 non-zero entries. 
In contrast, for matrices with larger NNZ, GPU methods benefit from their massive parallelism, often achieving higher effective memory bandwidth and throughput. Nonetheless, the peak GFLOPS (the highest point in the scatter plot) is achieved by LOOPS, as shown in Figure~\ref{fig:gflops-fp16}. This advantage is primarily attributed to its sustained high compute throughput, particularly for matrices with high block density. 
For example, on the \texttt{TSOPF\_RS\_b2383} matrix, where each block contains on average 25.10 non-zeros out of a maximum of 32, LOOPS achieves 1964.9 GFLOPS, substantially outperforming cuSparse (546.6 GFLOPS) and Magicube (336.9 GFLOPS).

Notably, the performance advantage of LOOPS becomes increasingly prominent at lower precisions, aligning with the design characteristics of SME, which provides significantly higher compute throughput for low-precision operations. Furthermore, for large matrices with regular sparsity patterns, LOOPS can achieve remarkably high performance that surpasses that of GPU methods, highlighting the advantage of structured computation and effective hardware utilization on Armv9 CPUs.

\subsection{Effectiveness of adaptive scheduling}
\label{sec:evaluation:scheduing}
We evaluate the effectiveness of our adaptive scheduling strategy in LOOPS by comparing it against two baselines: a pure NEON SpMM ($r_{boundary}=r_{total}$) implementation and a pure SME SpMM implementation ($r_{boundary}=0$). The goal is to assess how well our approach leverages the NEON and SME units simultaneously. Experimental results on the entire SuiteSparse Matrix Collection demonstrate that LOOPS achieves the best performance on 83.3\% of the matrices. On average, LOOPS delivers a 45.64$\times$ speedup over the pure NEON SpMM and a 124.72$\times$ speedup over the pure SME SpMM. 

The substantial improvement over the SME-only baseline is primarily attributed to the inefficiency of multi-threaded SME execution for many matrices. Since SME resources are limited and shared across cores, aggressive parallel execution often leads to resource contention and degraded performance. In contrast, the adaptive scheduling of LOOPS dynamically selects single-threaded SME execution when appropriate, effectively avoiding contention and delivering superior performance for such workloads. The reason why LOOPS cannot deliver the best performance in some cases is that those matrices with extremely small row sizes, high sparsity irregularity, or very low computation-to-communication ratios tend to cause load imbalance or under-utilization of either NEON or SME units. Future work will enhance LOOPS to dynamically identify such corner cases at runtime and adapt the scheduling policies accordingly.

\subsection{Energy efficiency}
\label{sec:evaluation:power}
To evaluate the energy efficiency of our proposed method on Apple M4Pro, we compare the power consumption, runtime, and energy efficiency (GFLOPS/W) of LOOPS against cuSparse on NVIDIA A100 for 20 representative matrices in FP16 precision. 
The power consumption is measured by \texttt{powermetrics} tool and NVIDIA Management Library (NVML)~\cite{NVML} on Apple M4Pro and NVIDIA A100 respectively.
It is worth noting that the number of the NNZ of all these 20 matrices exceeds $10^6$ which can benefit from GPU execution more.
As shown in Table~\ref{tab:a100_m4pro}, although the NVIDIA A100 achieves shorter runtime for most matrices due to its massive parallel throughput, it consumes significantly more power, averaging over 240W, compared to around 10--40W on Apple M4Pro. Interestingly, for several matrices such as \texttt{m6}, \texttt{m8}, and \texttt{m14}, and \texttt{m17}, LOOPS not only demonstrates lower power consumption but also achieves faster runtime than the cuSparse, resulting in up to 20$\times$ energy usage reduction.
Overall, the results clearly indicate that LOOPS provides excellent energy efficiency. The benefits become more prominent in workloads with limited compute intensity, where high-end GPUs are often underutilized but still consume considerable power.

\begin{table}[htbp]
\centering
\footnotesize
\caption{Power, time, and efficiency comparison between cuSparse on NVIDIA A100 and our method on Apple M4Pro for 20 matrices in FP16 precision.}
\label{tab:a100_m4pro}
\begin{tabular}{l|ccc|ccc}
\hline
\multirow{3}{*}{Id} & \multicolumn{3}{c|}{NVIDIA A100} & \multicolumn{3}{c}{Apple M4Pro} \\
 & Power & Time & Eff. & Power & Time & Eff. \\
 & (W) & (s) & (GFLOPS/W) & (W) & (s) & (GFLOPS/W) \\
\hline
m1 & 247.8 & \textbf{0.8910} & 1.73 & \textbf{38.4} & 2.603 & \textbf{3.81} \\
m2 & 245.8 & \textbf{0.1978} & 1.98 & \textbf{6.3} & 0.3678 & \textbf{41.46} \\
m3 & 230.6 & \textbf{0.2770} & 1.85 & \textbf{36.9} & 0.5105 & \textbf{6.28} \\
m4 & 247.4 & \textbf{0.1979} & 2.21& \textbf{9.5} & 0.3771 & \textbf{30.22} \\
m5 & 242.3 & \textbf{0.2667} & 1.91 & \textbf{10.2} & 0.4350 & \textbf{27.75} \\
m6 & 247.3 & 0.1307 & 2.30 & \textbf{29.9} & \textbf{0.1079} & \textbf{23.08} \\
m7 & 247.5 & \textbf{0.3967} & 1.74 & \textbf{31.0} & 1.190 & \textbf{4.62} \\
m8 & 248.3 & 0.0929 & 2.87 & \textbf{12.3} & \textbf{0.0636} & \textbf{84.70} \\
m9 & 235.9 & \textbf{0.0411} & 1.39 & \textbf{38.1} & 0.1136 & \textbf{3.11} \\
m10 & 243.4 & \textbf{0.0599} & 1.36 & \textbf{37.8} & 0.1902 & \textbf{2.76} \\
m11 & 242.2 & \textbf{0.0889} & 2.32 & \textbf{30.8} & 0.0985 & \textbf{16.48} \\
m12 & 224.6 & \textbf{0.0233} & 1.56 & \textbf{41.8} & 0.1162 & \textbf{1.68} \\
m13 & 211.4 & \textbf{0.0171} & 1.70 & \textbf{42.7} & 0.0562 & \textbf{2.56} \\
m14 & 236.3 & 0.0438 & 2.69 & \textbf{10.2} & \textbf{0.0382} & \textbf{71.36} \\
m15 & 247.2 & \textbf{0.0663} & 2.35 & \textbf{14.3} & 0.0744 & \textbf{36.16} \\
m16 & 240.6 & \textbf{0.0440} & 2.42 & \textbf{17.7} & 0.0640 & \textbf{22.64} \\
m17 & 96.5 & 0.2023 & 0.86 & \textbf{16.5} & \textbf{0.1194} & \textbf{8.53} \\
m18 & 202.1 & \textbf{0.0133} & 1.82 & \textbf{33.6} & 0.0537 & \textbf{2.72} \\
m19 & 236.2 & \textbf{0.0235} & 2.74 & \textbf{17.3} & 0.0431 & \textbf{20.38} \\
m20 & 247.7 & \textbf{0.0675} & 0.89 & \textbf{8.4} & 0.1413 & \textbf{12.56} \\
\hline
\end{tabular}
\end{table}

\subsection{Case study: End-to-end GNNs performance}
\label{sec:evaluation:dgl}
\begin{table}[htbp]
\centering
\small
\caption{Statistics of the evaluated datasets.}
\renewcommand{\arraystretch}{1.2}
\begin{tabular}{>{\centering\arraybackslash}m{1.5cm} | >{\centering\arraybackslash}m{3cm} | >{\centering\arraybackslash}m{1cm} | >{\centering\arraybackslash}m{1.5cm}}
\hline
\textbf{Dataset} & \textbf{Description} & \textbf{\#Nodes} & \textbf{\#Edges} \\
\hline
Reddit & Reddit posts from September 2014 & 232,965 & 114,615,892 \\
\hline
Amazon-ratings & Amazon co-purchasing network & 24,492 & 186,100 \\
\hline
Yelp & Yelp business network & 716,847 & 13,954,819 \\
\hline
\end{tabular}
\label{tab:dataset_statistics}
\end{table}
We integrate LOOPS into DGL~\cite{wang2019deep} framework and choose a popular GNN model, specifically a Graph Convolutional Network (GCN)~\cite{gibert2012graph,kipf2016semi} for end-to-end evaluation. GNN models mainly consist feature aggregation and feature update processes. For GCN, it aggregates features of neighbor nodes by the SpMM operator. As shown in Table~\ref{tab:dataset_statistics}, the graph datasets are selected from the built-in datasets of DGL, which is popular for evaluating the performance of GNNs. 
The end-to-end time includes the time of format conversion and model training. We select the latest DGL 2.5 with libxsmm~\cite{heinecke2016libxsmm} backend enabled for performance comparison. Compared with DGL, LOOPS achieves speedups of 2.81$\times$, 1.08$\times$, and 1.12$\times$ for the three graphs. The difference in speedups is attributed to the block density. Under LOOPS format, the block density of Reddit dataset is 2.068, while the block density of Amazon-ratings dataset is only 1.028. This indicates that our method can bring more performance benefits when the nonzero entries of the dataset are more gathered.
Meanwhile, the training accuracy of GCN using LOOPS kernel is consistent with the original DGL (no accuracy loss). Additionally, for static GNN scenarios, preprocessing only needs to be performed once and can be easily amortized by multiple training epochs. Specifically, the preprocessing
overhead accounts for only 1.3\% of the end-to-end GNN traning time. 

%% file: 5-relatedwork.tex
\section{Related Work}
\label{sec:relatedwork}

\textbf{CPU SpMM.}  
Optimized SpMM kernels on general-purpose CPUs remain relatively underexplored. Several existing efforts such as FusedMM~\cite{rahman2021fusedmm}, FeatGraph~\cite{hu2020featgraph}, and Rosko~\cite{natesh2023rosko} focus on using SIMD instructions to accelerate SpMM. Additionally, vendor-provided libraries for CPUs, such as Intel MKL~\cite{wang2014intel} or ArmPL~\cite{ArmPL2504}, provide strong support for sparse linear algebra. However, ArmPL supports SpMV and SpGEMM, but does not support SpMM. Code-generation frameworks and auto-tuning systems such as TVM~\cite{chen2018tvm}, LIBXSMM~\cite{heinecke2016libxsmm} and TACO~\cite{kjolstad2017tensor} aim to generate optimized kernels for sparse workloads. However, they often struggle to adapt to rapidly evolving architectural features such as Armv9’s SME. For example, LIBXSMM currently provides SME support only for dense GEMM kernels. Moreover, auto-tuning frameworks require significant search time and may introduce considerable tuning overhead. 

\textbf{GPU SpMM.}  
Extensive work has been devoted to optimizing SpMM on GPUs. Sputnik~\cite{gale2020sparse} presents the one-dimensional tiling scheme and leveraging CUDA cores to accelerate sparse computation. vectorSparse~\cite{chen2021efficient} proposes the TCU-based 1-D Octet tiling method, using vectorized memory access. GE-SpMM~\cite{huang2020ge} employs the warp merging strategy to reduce redundant data loading. To improve load balancing, RoDe~\cite{pang2024row} proposes a 2D strategy to divide the rows of sparse matrices into regular and residue parts and introduces new load balance and fine-grained pipeline technologies for further optimization. 
Current SOTA works typically use TCUs to push the SpMM performance to a new height. TC-GNN~\cite{wang2023tc} and DTC-SpMM~\cite{fan2024dtc} propose custom sparse formats which divides the sparse matrix into nonzero vectors for computation on the TCUs. Magicube~\cite{li2022efficient} is a high-performance
sparse-matrix library to accelerate sparse and quantized matrix operations in deep learning leveraging the Strided Row-major BCRS (SR-BCRS) format. These approaches reconcile the sparse computation with the high-performance TCUs and employ multiple optimizations, including custom compressed format, reordering, load balancing, and pipeline optimizations. 

\textbf{DSA SpMM.}  
Numerous domain-specific accelerators \cite{hegde2019extensor,song2022sextans,gerogiannis2023spade,gerogiannis2024hottiles} have been proposed to address the challenges of sparse computation. These accelerators primarily exploit sparsity patterns using customized memory hierarchies and compute units. For instance, Sextans~\cite{song2022sextans} is a high performance FPGA accelerator for SpMM, which enables fast random access using on-chip memory, streaming access to off-chip large matrices, as well as PE-aware non-zero scheduling for balanced workload. SPADE~\cite{gerogiannis2023spade} is a hardware accelerator for SpMM and SDDMM which avoids data transfer overheads by tightly coupling accelerator PEs with the cores of a multicore and attains flexibility and programmability by supporting a tile-oriented ISA. HotTiles~\cite{gerogiannis2024hottiles} features heterogeneous SpMM accelerator architectures, which employ different types of PEs to exploit the insight that nonzeros form dense and sparse regions, rather than being uniformly distributed across the whole matrix. HYTE~\cite{li2025hyte} is a hybrid static-dynamic framework to enable flexible and efficient tiling on sparse accelerators and can flexibly adapt to the specific data sparsity patterns. While such solutions demonstrate high performance and efficiency, they often lack flexibility and are not readily deployable on general-purpose processors.

Different from these works, LOOPS targets SME and NEON units on modern Arm architectures and introduces a hybrid sparse format and performance model to fully exploit the heterogeneity in compute resources. Additionally, LOOPS achieves better (in small scale matrices) or comparable performance (in large scale matrices) against GPU methods with much lower power consumptions for SpMM on CPU.

%% file: 6-conclusion.tex
\section{Conclusion}
\label{sec:conclusion}
In this paper, we propose LOOPS, a novel framework that integrates a hybrid sparse matrix format with a coordinated execution strategy to enable efficient collaboration between NEON and SME units on Armv9-based CPUs. LOOPS combines the CSR and BCSR layouts to map different portions of the workload to the distinct hardware execution units. Additionally, LOOPS supports multi-precision SpMM (FP64/FP32/ FP16) and incorporates a lightweight performance model to guide dynamic scheduling decisions, achieving high performance and energy efficiency on the Apple M4Pro chip. Extensive experimental results show that LOOPS delivers state-of-the-art performance and energy efficiency for SpMM as well as end-to-end GNN performance.